# Design of Millimeter-wave Detector for Gyrotron Power Monitoring

Weiye Xu[1, a)], Handong Xu[1, b)], Fukun Liu[1], Xiaojie Wang[1]

[1] *Institute of Plasma Physics, Chinese Academy of Sciences, Hefei, Anhui, China*

a) b) Corresponding authors: xuweiye@ipp.ac.cn, xhd@ipp.ac.cn

**Abstract.** The real-time power monitoring of gyrotron is one of the key issues in the operation of electron cyclotron resonance heating system. The detector can be used for real-time power monitoring. We analyzed the principle of diode detection and designed a D-band wideband detector based on Schottky diode in this paper. The detector includes a waveguide-to-microstrip transition, a matching circuit, a diode, and a low pass filter. A novel waveguide-to-microstrip transition was developed based on probe coupling. A wideband lossy matching circuit was developed based on tapered-line and series matching resistor. The simulation results show that when the input power is -30dBm at 140 GHz, the detection sensitivity is about 1600V/W.

## INTRODUCTION

Electron cyclotron resonance heating (ECRH) [1] is one of the important plasma-assisted heating methods in controlled magnetic confinement fusion research. The RF power measurement of the gyrotron [2] is one of the key issues in the operation of electron cyclotron resonance heating system. There are two methods for measuring gyrotron power, namely the calorimetric method [3] and the detector-based method for real-time power monitoring [4]. For the detector-based method, the power monitor bends are used to leak out a little incident RF power and reflected RF power to the detectors, which will transform the wave power into voltage signal according to the sensitivity curve of the detector.

Diodes are widely used in detector designs. For example, H. Kazemi et al. proposed a new ErAs/InAlGaAs Schottky diode detector [5]. H. Meinel proposed a finline detection circuit that uses a zero-biased Schottky diode for wide-band, high-sensitivity detection [6]. There are commercial detectors based on finline detection in the 18 GHz to 170 GHz band at Farran, Mi-Wave, etc. In order to improve performance (the sensitivity, the signal-to-noise ratio, etc.), the balanced detectors using dual diodes have also been studied [7]. Transistors can also be used in detector designs. For example, V. Vassilev et al. proposed the 140-220 GHz detectors based on a 250-nm InP–InGaAs–InP double heterojunction bipolar transistor process [8]. Millimeter wave detectors using differential detection circuits have also been studied. For example, Le Zheng et al. reported a W-band detector based on a 0.18 μm SiGe BiCMOS [9] with a detection sensitivity of up to 91 V/mW. For millimeter wave frequency detection, Schottky diodes are more often used, and the cutoff frequency of the diode is now up to THz.

Considering the performance requirements and the processing difficulties of the detector for gyrotron power monitoring in EAST ECRH system, we designed a single diode based D-band (110-170GHz) detector. The detector consists of a waveguide-to-microstrip transition, a matching circuit, a detector diode, and a low-pass filter, which is shown in Fig. 1. The diode is more suitable for transmitting signals using the microstrip line, the microstrip based low pass filter and matching circuit are also easier to implement. However, the waveguide is more suitable for transmitting millimeter waves. It is necessary to couple the microwaves from the waveguide into the microstrip line. A novel waveguide-to-microstrip transition [10] was developed based on probe coupling. In this paper, according to the design process, the detector diode, low-pass filter, and matching circuit are discussed in turn, and then the designed detector is simulated and analyzed.

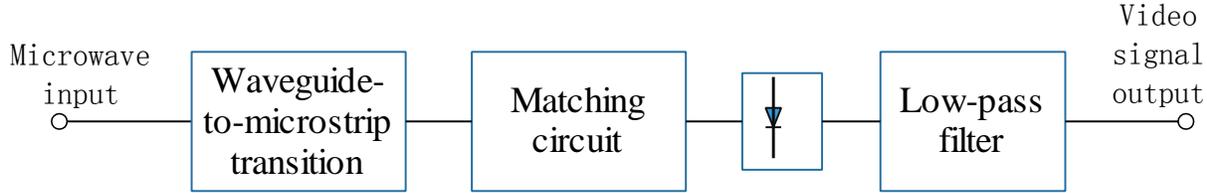

FIGURE 1. Diagram of detection circuit

## MILLIMETER WAVE DETECTOR DIODE

The ideal ampere-volt (I-V) characteristic of the diode is [11],

$$i(V) = I_s \left(e^{\frac{qV}{kT}} - 1\right) \quad (1)$$

Where $I_s$ is the saturation current of the diode under zero bias. For Schottky diodes, $I_s$ is generally in the range of $10^{-14} \sim 10^{-6}$ A, for example, the saturation current under zero bias of the W-band Schottky diode from VDI is 4~24 μA, and the saturation current of TSC-SB-03020 Schottky diode from TELEDYNE is $10^{-14}$ A; q is the electron charge; $V$ is the voltage between the anode and the cathode of the diode; k is the Boltzmann constant $1.38 \times 10^{-23}$ $J/K$; $T$ is the junction temperature in K.

In fact, the diode has parasitic effects such as series resistance effect, and the actual operating characteristic curve can be divided into three parts: the cut-off area, the working area and the saturation area [12]. The V-I characteristic in the working area is,

$$i(V) = I_s \left(e^{\frac{qV}{nkT}} - 1\right) \quad (2)$$

Where $n$ is an ideal factor, ideally, its value is 1. Due to the influence of junction defects, etc., the value of $n$ is greater than 1, for surface contact Schottky diodes, $n$ is between 1 and 1.2; for point contact Schottky diodes, $n$ is generally greater than 1.4.

Suppose $V_p$ is the amplitude of the high-frequency signal. For small signal conditions, $\frac{qV_p}{nkT} \ll 1$, the DC component of the diode output signal is,

$$V_{DC} = \frac{nkT}{q}\left(\frac{qV_p}{2nkT}\right)^2 = \frac{qV_p^2}{4nkT} \quad (3)$$

As can be seen from the above equation, the detector works in the square rate law. The output voltage is proportional to power.

For large signal conditions, that is, the amplitude of the input high-frequency signal is greater than the diode turn-on voltage,

$$V_{DC} \approx V_p \quad (4)$$

The diode can be analyzed with an equivalent circuit, which is shown in Fig. 2. Wherein, Ls is a series inductance of the pin; $R_s$ is a series resistance caused by contact, diffusion resistance, etc.; $R_j$ is a junction resistance, the resistance is related to the bias voltage of the diode; $C_j$ is a diode junction capacitance, and the capacitance value is related to the bias voltage of the diode;

$C_{pp}$ is a diode shell capacitor (capacitance between two pins - Pad to Pad Capacitance). The cutoff frequency of the Schottky diode is,

$$f_c = \frac{1}{2\pi R_s C_j} \tag{5}$$

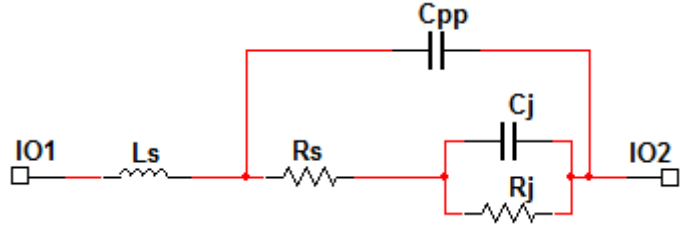

FIGURE 2. The equivalent model of the diode.

Generally, the turn-on voltage of a non-zero bias Schottky diode is about 0.4V; the turn-on voltage of a zero-biased Schottky diode (ZBD) is generally less than 100mV. When a ZBD is applied to a detection circuit, it is not necessary to add a bias voltage, which reduces the complexity of the system. However, the I-V characteristics of a ZBD are not ideal for non-ZBD.

There are many manufacturers that produce millimeter-wave or THz Schottky diodes, such as Keysight (split from Agilent in 2013), VIRGINIA Diodes Inc. (VDI, USA), Teratech (split from Rutherford Laboratories, UK in 2010), United Monolithic Semiconductors (France), Teledyne Scientific & Imaging (USA). We chose to use a G-band (110-300GHz) ZBD produced by VDI. The minimum chip dimensions are 255×88×43 μm (length×width×thickness). The typical forward I-V curve of VDI G-band ZBD is shown in Fig. 3.

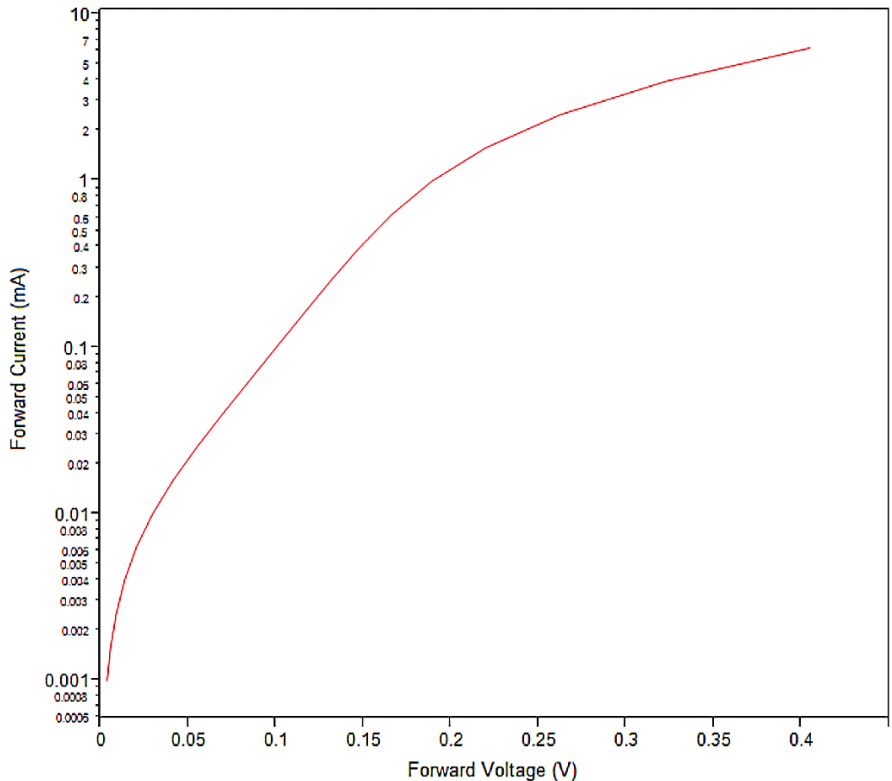

FIGURE 3. Typical forward I-V curve of VDI G-band ZBD.

The equivalent circuit of VDI G-band ZBD is shown in Fig. 4. Where, $C_j \approx C_T - C_{PP} = 12$ fF-7 fF=5 fF. When the video resistor $R_v$ is biased at 0V, it can be obtained by dividing the voltage value near 0V by the current value, and its value varies with temperature, which is about 2~7.5 kΩ. We take four points from Fig. 3 ([voltage in mV, current in mA]: [200, 1], [150, 0.5], [70, 0.04], [50, 0.02]) to calculate $R_s$,

$$R_s = \frac{(V_4 - V_3) - (V_2 - V_1)}{(I_4 - I_3) - (I_2 - I_1)} = \frac{(200 - 150) - (70 - 50)}{(1 - 0.5) - (0.04 - 0.02)} \approx 62.5 \, \Omega \tag{6}$$

So the zero bias junction resistance is,

$$R_j = R_v - R_s \approx R_v \tag{7}$$

Since the junction resistance and saturation current have the following relationship [13],

$$R_j = \frac{dV}{dI} = \frac{V_o}{I_{sat}} \tag{8}$$

where $V_o = nkT/q$, the ideality factor is,

$$n = \frac{q(V_2 - V_1)}{kT \ln(I_2/I_1)} = \frac{q(70-50)[mV]}{kT \ln 2} \approx 1.12 \tag{9}$$

The saturation current is,

$$I_{sat} = \frac{V_o}{R_j} = \frac{nkT}{qR_j} \approx \frac{0.029}{3500} \approx 8.4 \, \mu A \tag{10}$$

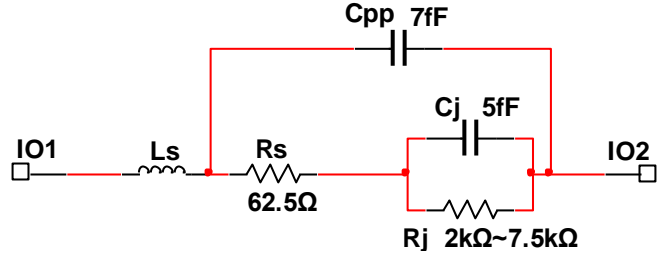

FIGURE 4. Equivalent circuit of VDI G-band ZBD.

## LOW-PASS FILTER

The low pass filter is used to filter out high-frequency signals and only output low-frequency signals related to power. There are many types of microstrip filters [14], such as interdigital filters, parallel coupled microstrip line filters, high and low impedance line filters, microstrip defected ground structure filters [15] and the like.

When the millimeter wave is input into the detection diode, the DC/low frequency and harmonic signals are generated. What we need is a DC/low-frequency signal proportional to the wave power. The fundamental wave and harmonics need to be suppressed. We only need to consider suppressing the fundamental wave and the second harmonics, because the amplitude of the higher harmonic signal is very small, which can be neglected compared to the low-frequency signal.

First, the filter prototype consisting of the inductor and capacitor lumped components is calculated according to the filter design target, and then the inductors and capacitors are converted into high-low impedance lines, radial stubs, short-circuited branches, open-circuited branches, etc.

We want the filter circuit to take up as little space as possible and choose a low-pass filter based on a butterfly radial stub. We designed the filter passband (3dB) with an upper frequency limit of 50GHz and a 20dB cutoff frequency of 100GHz. The fourth-order maximum flattening filter is used to obtain the lumped parameter filter and its amplitude-frequency response as shown in the two figures below.

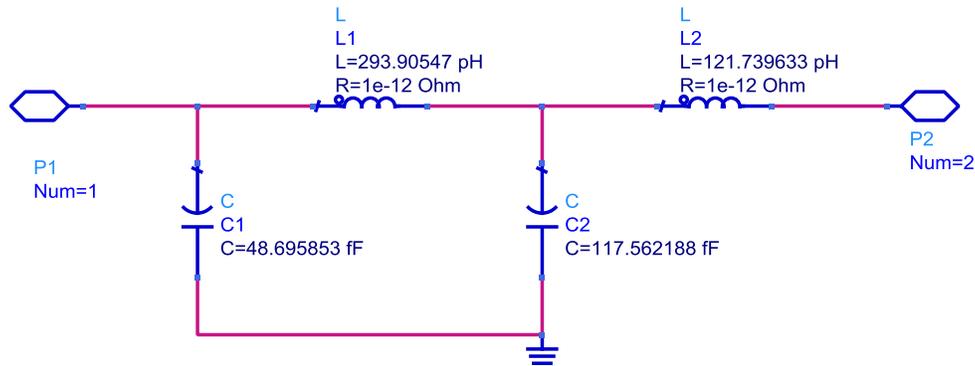

FIGURE 5. Low-pass filter lumped parameter model with a 20dB cutoff frequency of 100GHz.

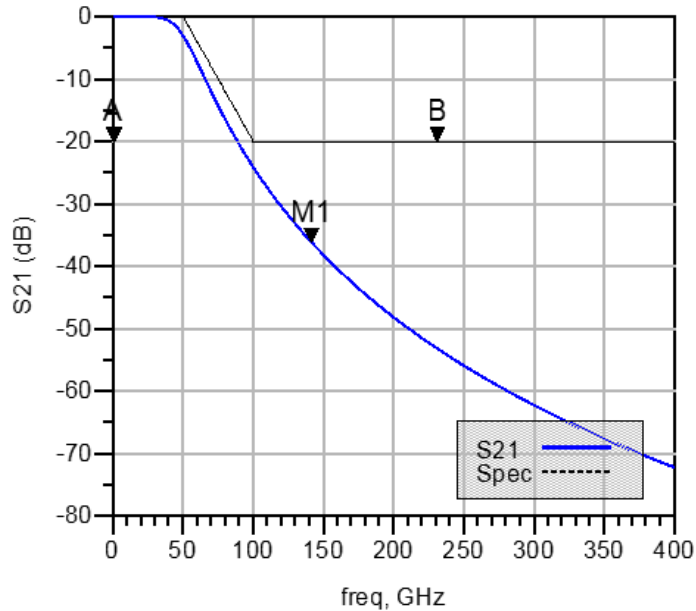

FIGURE 6. The amplitude-frequency response of the designed fourth-order low-pass filter.

Then, we convert the lumped elements into microstrip lines. There are many empirical formulas that can be used. The ADS can be used to easily obtain the corresponding microstrip circuits. The substrate materials suitable for the D-band (110~170 GHz) are alumina ceramics, sapphire, RT_Duriod5880, etc. Considering factors such as cost and performance, we chose RT_Duriod5880 whose dielectric constant is 2.2 as the substrate. The dielectric thickness is designed to be 5 mils, the microstrip line material is copper, and the microstrip line thickness is 0.7 mil. We convert the lumped component capacitance into a butterfly radial stub, and to convert the lumped inductance into a high impedance line. The high-impedance line impedance is 100 ohms, the corresponding microstrip line width is 4.494 mils, and the lengths of the two segments are 26.1 mils and 10.8 mils, respectively. As shown in Fig. 7 and Fig. 8, the two butterfly radial stubs have a W of 15.6 mils, an angle of 60°, and a radius R0 of 13.8 mils and 20.7 mils, respectively. The simulation results shown in Fig. 9 show that the insertion loss is 32.8dB and 21dB respectively at 140GHz and 280GHz, and the harmonic signals are better suppressed.

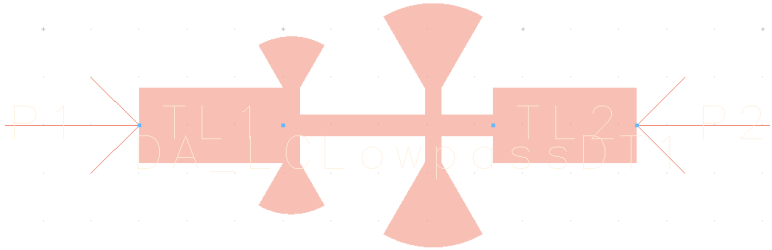 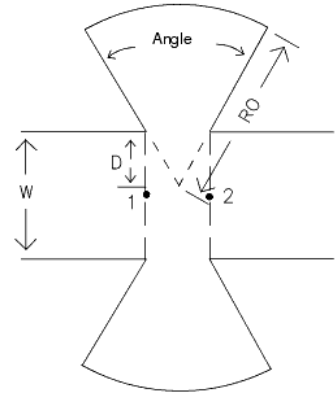

FIGURE 7. The microstrip circuit of the low pass filter.   FIGURE 8. The structure of the butterfly radial stub.

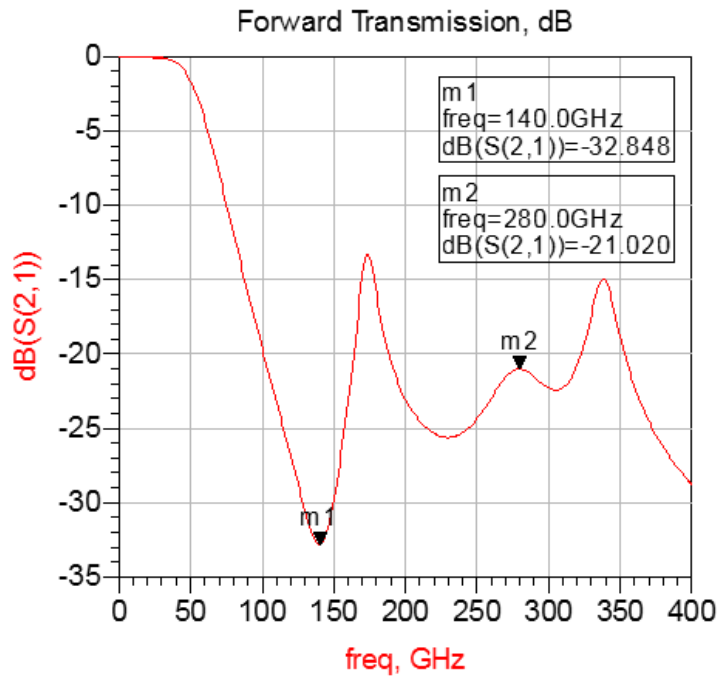

FIGURE 9. The ADS simulation results of the low pass filter.

In order to further verify the performance of the filter, we imported the PCB (printed circuit board) layout into the EMPro 3D electromagnetic simulation software and simulated it with the finite element method (FEM). The simulation results show that the insertion loss is less than -40dB in the 140GHz and 280GHz bands.

## MATCHING CIRCUIT

The matching [16] circuit is used to match the input impedance of the detector to a 50 Ω impedance to reduce the reflection, so that the microwave power enters the diode as much as possible, improving the detection sensitivity. There are many implementations of the microstrip line matching circuits, such as branch matching, tapered-line matching, and fan-shaped stub matching. The matching circuit can easily be designed using the Smith chart.

Since the diode IO2 port (shown in Fig. 2) is connected to the low-pass filter, the input impedance of the designed low-pass filter is first analyzed. The simulation results are shown in Fig. 10, the impedance is 1.56-j*22.5 at 140 GHz, is 26.1-j*230.5 at 110 GHz, is 34.5-j*60.7 at 170 GHz.

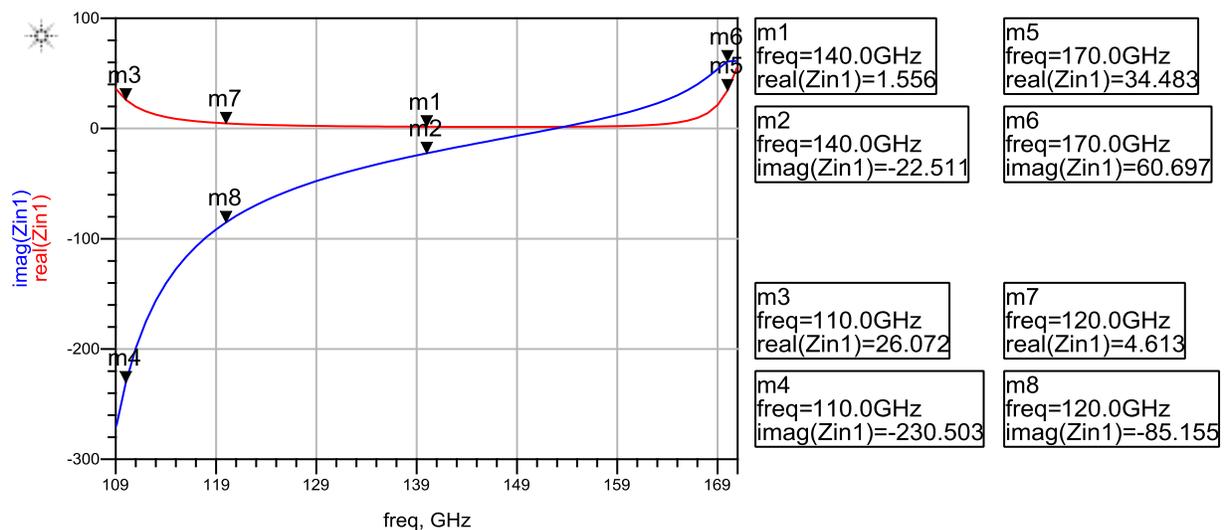

FIGURE 10. The input impedance of the low pass filter.

The value of $L_s$ in the diode equivalent circuit shown in Fig. 4 is unclear. It is estimated that the value of $L_s$ is less than 0.05 nH. We first ignore $L_s$ to analyze the diode input impedance. The diode input impedance analysis circuit is shown in Fig. 11. First, we set the diode output resistance to 1.56-j*22.5 (the low-pass filter input impedance at 140 GHz). The parameter sweep of $R_j$ is performed with a step-size of 1.5kΩ (from 2kΩ to 7.5kΩ). The simulation results shown in Fig. 12 show that the value of $R_j$ has little effect on the impedance. At 140 GHz, when Rj is 2kΩ, the diode input impedance is 16.3-j*120.1, when $R_j$ is 3.5kΩ, it is 14.5-j*119.6, and when $R_j$ is 7.5kΩ, it is 13.2-j*119.3. And the impedance varies with frequency, but the change is not very large. Then we set the diode output impedance to the low-pass filter input impedance of 26.1-j*230.5 at 110 GHz. The input impedance of the diode is 40.7-j*353.3 at 110 GHz when $R_j$ is 3.5 kΩ, which also does not change much with frequency. When the output impedance of the diode is set to a low-pass filter input impedance of 34.5-j*60.7 at 170 GHz, and the input impedance of the diode is 46.5-j*141.3 at a frequency of 170 GHz when $R_j$ is 3.5 kΩ, which also has little change with frequency.

In order to further confirm the above simulation results, we built a diode simulation model for VDI G-Band ZBD in ADS, which is shown in Fig. 13. After replacing the diode equivalent circuit in Fig. 11 with the diode model shown in Fig. 13, the simulation results are similar.

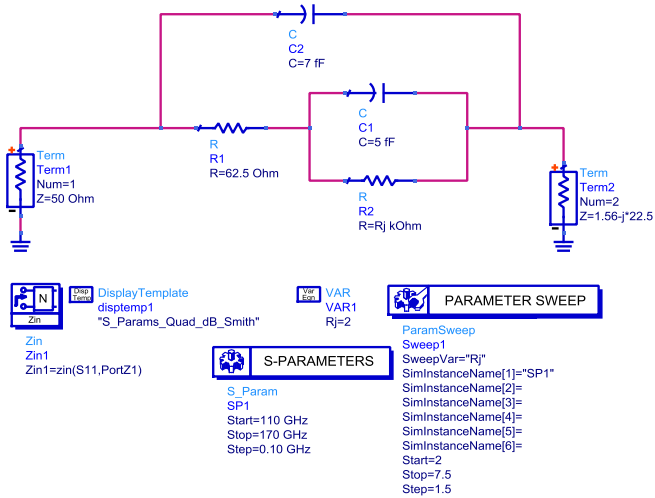

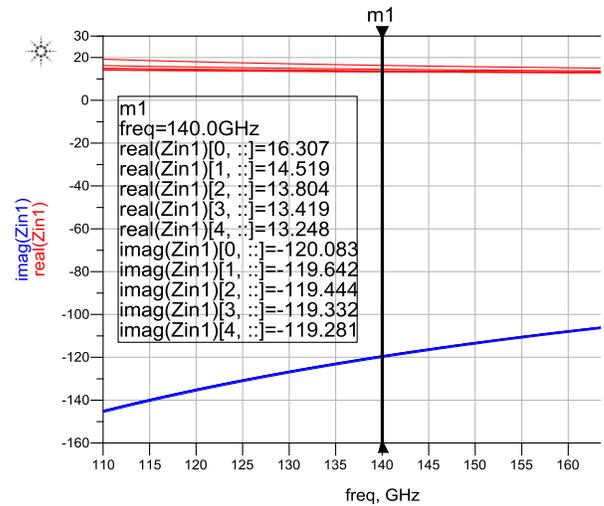

FIGURE 11. Diode input impedance analysis circuit.

FIGURE 12. Diode input impedance simulation results at 140 GHz.

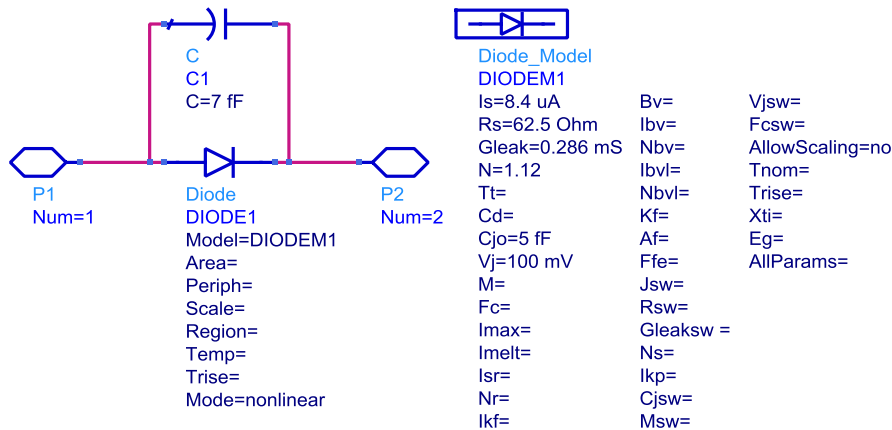

FIGURE 13. Diode model of VDI G-band ZBD.

Since the gyrotron output frequency in the EAST ECRH system is 140 GHz and $R_j$ of the VDI ZBD is about 3.5kΩ at normal temperature, the input impedance of the diode should be matched to 14.5-j*119.6 Ω. The input impedance of the diode varies with frequency. In order to make the diode have the highest sensitivity in the 110~170GHz band, a broadband impedance matching method combining a tapered-line [17] match and a series resistor match is proposed. First, a resistor is connected in series with the diode input to match the impedance of 14.5-j*119.6 to 14.5 Ω. The larger the selected resistance value, the wider the available bandwidth, but the lower the sensitivity. We hope that the detection sensitivity is greater than 1000 at 140 GHz, and finally, R=8.6 Ω is selected. Then a tapered-line of 14.5 Ω to 50 Ω is used to obtain the final matching circuit, which is shown in Fig. 14 and Fig. 15. The simulation result of the matching circuit when load impedance is set to 14.5-j*119.6 Ω is shown in Fig. 16. The VSWR is 1.18 at 140 GHz.

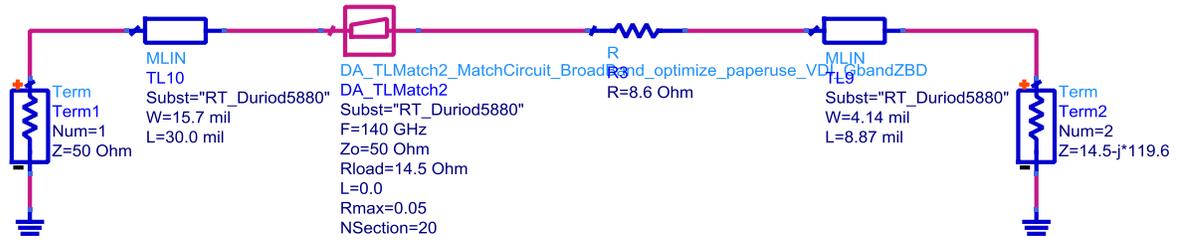

FIGURE 14. Schematic of the impedance matching circuit.

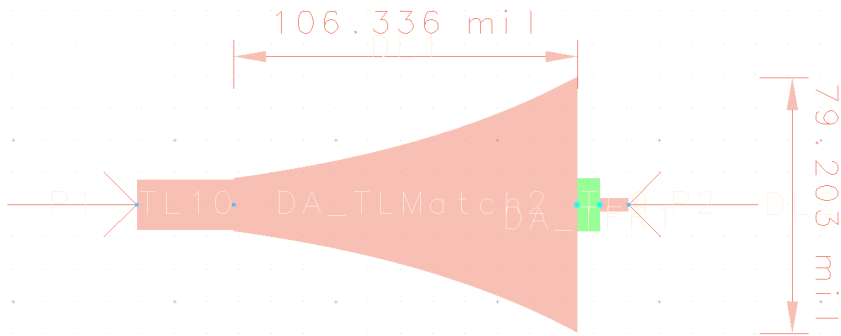

FIGURE 15. The layout of impedance matching circuit.

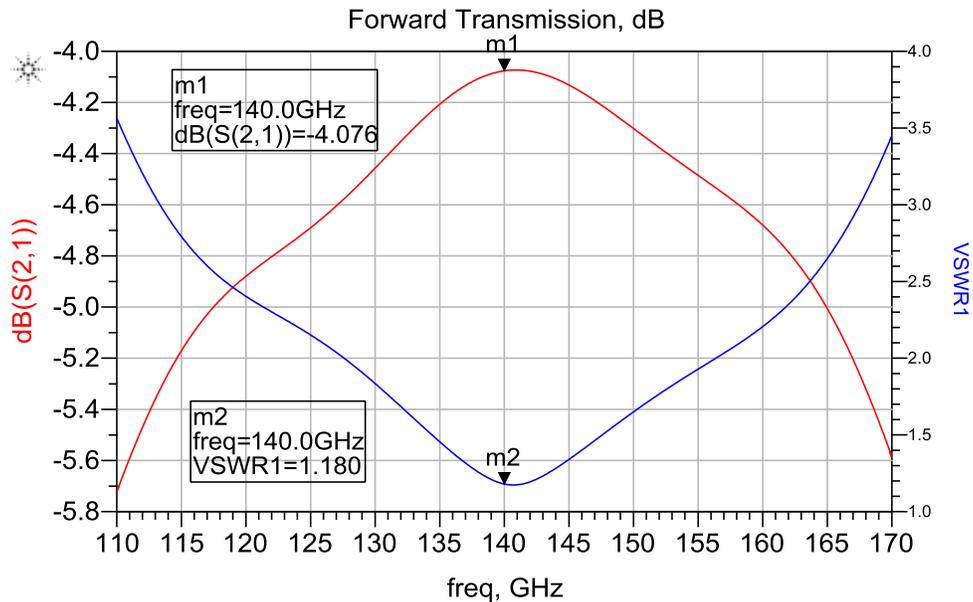

FIGURE 16. Simulation result of matching circuit when load impedance is set to 14.5-j*119.6 Ω.

We simulated the matching circuit in the 110GHz~170GHz band. The results are shown in Table 1. The results show that the VSWR in the 120GHz~170GHz band is less than 8.26, and the VSWR in the full range of 110GHz~170GHz is less than 22.52.

TABLE 1.

| Frequency [GHz] | Input impedance of low pass filter [Ω] | Input impedance of the diode when $R_j=3.5$ kΩ [Ω] | S21 [dB] | VSWR | Power delivered to the diode when the input power is 1W [W] |
|---|---|---|---|---|---|
| 110 | 26.1-j*230.5 | 40.7-j*353.3 | -13.901 | 22.512 | 0.041 |
| 120 | 4.6-j*85.2 | 18.6-j*198.0 | -9.492 | 8.258 | 0.11 |
| 130 | 2.2-j*44.9 | 15.6-j*149.3 | -6.333 | 3.633 | 0.23 |
| 140 | 1.56-j*22.5 | 14.5-j*119.6 | -4.076 | 1.180 | 0.39 |
| 170 | 34.5-j*60.7 | 46.5-j*141.3 | -2.673 | 2.058 | 0.54 |

## DETECTOR AND THE SIMULATION RESULTS

The detector circuit consisting of a matching circuit, a diode, and a low-pass filter is shown in Fig. 17. The harmonic balance simulation of the detector is carried out. The simulation results (shown in Fig. 18) show that the detector has good linearity for a frequency of 140 GHz. When the input power is -10 dBm, the output voltage is about 0.108 V, and the detection sensitivity is about 1080 V/W. When the input power is -20 dBm, the output voltage is about 0.015V, the detection sensitivity is about 1500V/W; when the input power is -30dBm, the output voltage is about 0.0016V, and the detection sensitivity is about 1600V/W. The simulation results when L=0 are similar to the simulation results when L=0.05nH.

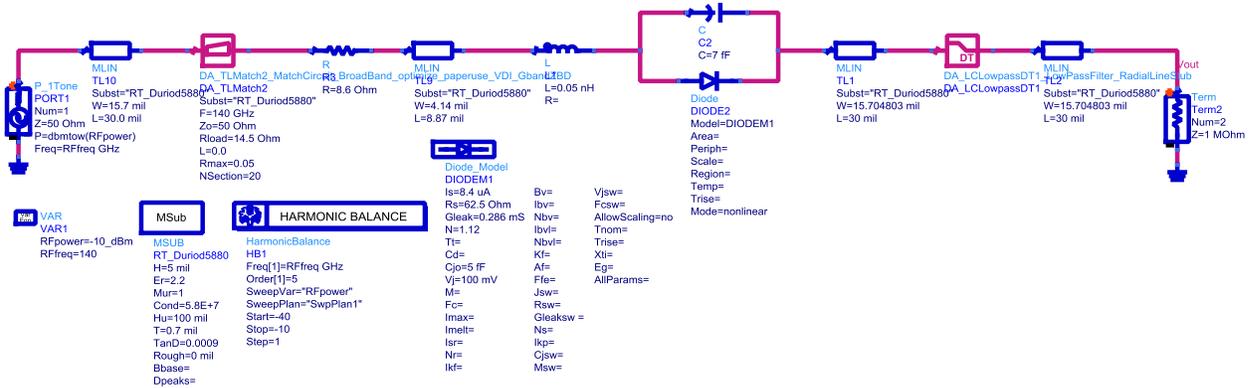

FIGURE 17. The detector circuit and harmonic balance simulation configuration.

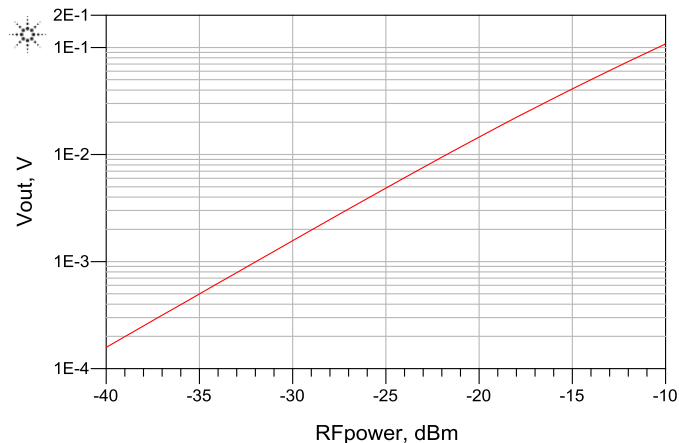

FIGURE 18. The harmonic balance simulation results at 140 GHz.

Change the sweep variable from RFpower to Rffreq. Set the series inductance to L=0. When the input power is set to -10dBm, the simulation results show that the detection sensitivity is greater than 500V/W in the frequency range of 123GHz~164GHz, which is shown in Fig. 19. When the input power is set to -20dBm, the simulation results show that the detection sensitivity is greater than 500V/W in the frequency range of 119GHz~166GHz. The series inductance (0 or 0.05nH) has little effect on the results. The sensitivity is close to (better than) that of the D-band Zero Bias Detector from Millitech, Inc. or Farran Technology over the entire D-band (at 140GH). The detector designed in this paper is more suitable for use in the EAST ECRH system.

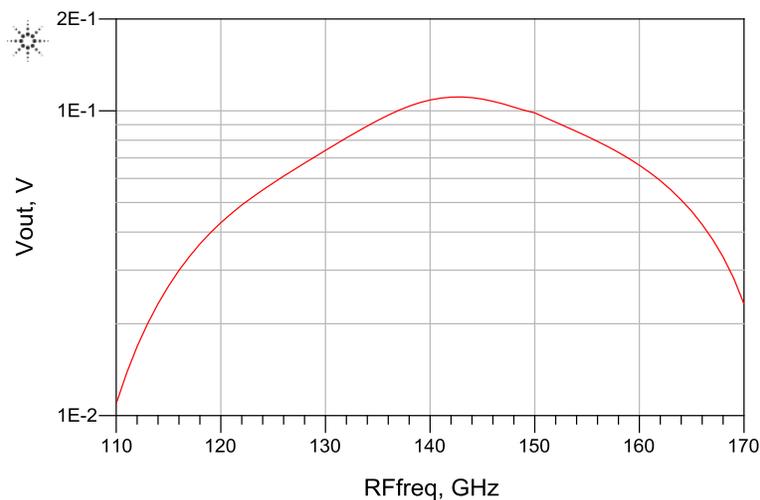

FIGURE 19. The harmonic balance simulation results at -10 dBm.

## SUMMARY

A D-band detector based on Schottky diode was designed in this paper. In order to increase bandwidth, a lossy matching circuit was developed based on tapered-lines and matching resistors. The VSWR is 1.18 at 140 GHz and is less than 8.26 in the 120GHz~170GHz band. The harmonic balance simulation results show that when the input power is -30dBm at 140 GHz, the detection sensitivity is about 1600V/W. The processing and testing of the detector will be done in the future.

## ACKNOWLEDGMENTS

This work was supported in part by the National Key R&D Program of China (grant no. 2017YFE0300401) and the National Magnetic Confinement Fusion Science Program of China (grant nos. 2015GB102003 and 2015GB103000).